# Observation of field induced anomalous quantum criticality in $Ce_{0.6}Y_{0.4}NiGe_2$ compound

Karan Singh and K. Mukherjee*

School of Basic Sciences, Indian Institute of Technology Mandi, Mandi 175005, Himachal Pradesh, India

E-mail: kaustav@iitmandi.ac.in

## ABSTRACT

We report the results of our investigation of magnetization and heat capacity on a series of compounds $Ce_{1-x}Y_xNiGe_2$ (x = 0.1, 0.2 and 0.4) under the influence of external magnetic field. Our studies of the thermodynamic quantity -d$M$/d$T$ on these compounds indicate that magnetic frustration persists in $Ce_{0.9}Y_{0.1}NiGe_2$, as also reported for the parent compound $CeNiGe_2$. The weak signature of this frustration is also noted in $Ce_{0.8}Y_{0.2}NiGe_2$, whereas, it is suppressed in $Ce_{0.6}Y_{0.4}NiGe_2$. Heat capacity studies on $Ce_{0.9}Y_{0.1}NiGe_2$ and $Ce_{0.8}Y_{0.2}NiGe_2$ indicate the presence of a new magnetic anomaly at high field which indicates that quantum criticality is absent in these compounds. However, for $Ce_{0.6}Y_{0.4}NiGe_2$ such an anomaly is not noted. For this later compound, the magnetic field ($H$) and temperature ($T$) dependence of heat capacity and magnetization obey $H/T$ scaling above critical fields. However, the obtained scaling critical parameter ($\delta$) is 1.6, which is away from mean field value of 3. This deviation suggests the presence of unusual fluctuations and anomalous quantum criticality in these compounds. This unusual fluctuation may arise from disorderness induced by Y-substitution.

## 1. Introduction

The low temperature properties of 4*f*-electron systems show a wide range of unusual magnetic and electronic behaviors, which is a broad and active field of research in the area of condensed matter physics over the past few decades [1].These systems are reported to exhibit various exciting phases, such as spin density wave, non-fermi liquid behavior, unconventional superconductivity, quantum criticality, antiferromagnetic and paramagnetic state etc. [2-6]. In some of these systems, the properties are dominated by the outcome of coupling between magnetic and elastic degree of freedom [7-8]. Also, observation of delocalization of some 4*f*-electrons along with the competition between the localized and delocalized moments provides an insight about the magnetic state of these systems [9 -11]. In some of the compounds belonging to this system, variation of external parameters like magnetic field, external and chemical pressure results in conspicuous discrepancies in the magnetic and thermal properties. In this context, $CeNiGe_2$ is an interesting 4*f*-electron based system, which has been studied in the last couple of decades and is also under investigation in recent years [12, 13]. It is also reported that in this compound external pressure and high magnetic fields are unable to suppress the long range ordering, which results in the absence of quantum critical point (QCP) [13-14]. However, Kim *et al*., reported that when Si is partially replaced by Ge, it leads to the suppression of magnetic ordering temperature and observation of some signatures of QCP [15]. Magnetic Gruneisen parameter ($\Gamma_{mag}$) and thermodynamic quantity (-d$M$/d$T$) are sensitive tools for the investigation of QCP [16]. In addition, $\Gamma_{mag}$ is used to identify magnetic frustrations and it displays a sign change in frustration regime at finite temperature [17, 18]. Our investigation on $CeNiGe_2$ through $\Gamma_{mag}$ revealed that there is a sign change in $\Gamma_{mag}$ from positive to negative, which is related to magnetic frustration in this compound. This partially frustrated regime develops a new antiferromagnetically ordered phase at high fields and field induced QCP is absent in this compound [19]. Y-substitution at the Ce-site results in sequential suppression of magnetic ordering temperature below 1.8 K. However, Non-Fermi liquid behavior or zero field QCP is absent even after 40% replacement of Ce [20]. Hence, it is of interest to study the magnetic field dependence properties of Ce-site diluted $CeNiGe_2$, to see the effect of magnetic frustration and possible quantum criticality.

With this objective, in this work, we report the evolution of magnetic and thermodynamic properties of $Ce_{1-x}Y_xNiGe_2$ (x = 0.1, 0.2 and 0.4) with magnetic fields. It is observed that the transition temperature decreases sequentially with the increase in the Y-concentration. The temperature response of (-d$M$/d$T$)/$T$ indicates that magnetic frustration is present in $Ce_{0.9}Y_{0.1}NiGe_2$; weakens in $Ce_{0.8}Y_{0.2}NiGe_2$, and is absent in $Ce_{0.6}Y_{0.4}NiGe_2$. Heat capacity studies indicate to that a new magnetic anomaly develops at high fields in $Ce_{0.9}Y_{0.1}NiGe_2$ and $Ce_{0.8}Y_{0.2}NiGe_2$ which avoids the quantum criticality in these compounds. However, such a new anomaly is absent in $Ce_{0.6}Y_{0.4}NiGe_2$. For this later compound, temperature and magnetic field dependent heat capacity and magnetization follow $H/T$ scaling. However, the obtained critical scaling parameter deviates from the mean field value. This observed deviation from the mean field value may be due to the presence of unusual fluctuations arising from disorderness induced by Y-substitution. This indicates the presence of anomalous quantum criticality in $Ce_{0.6}Y_{0.4}NiGe_2$ compound.

## 2. Experimental details

The polycrystalline compounds $Ce_{0.9}Y_{0.1}NiGe_2$ (Y-0.1), $Ce_{0.8}Y_{0.2}NiGe_2$ (Y-0.2) and $Ce_{0.6}Y_{0.4}NiGe_2$ (Y-0.4) are the same as used in Ref [20]. Temperature ($T$) dependent magnetization ($M$) is performed using Magnetic Property Measurement System (MPMS), while, temperature and magnetic field ($H$) dependent heat capacity ($C$) are performed using Physical Property Measurement System (PPMS); both from Quantum design, USA.

## 3. Results and discussion

Figure 1 (a, b and c) shows the temperature response of DC susceptibility ($M/H$) under zero field cooling condition at different magnetic fields (0.5 - 7 T) for Y-0.1, Y-0.2 and Y-0.4 compounds respectively. For Y-0.1 compound, a transition temperature ~ 3.3 K and a deviation ~2.2 K are observed which are shifted to lower temperature with increasing magnetic field and is suppressed below 1.8 K above 1 T. For the Y-0.2 compound, a transition temperature ~ 2.2 K is suppressed below 1.8 K at an applied field of 1 T whereas for the Y-0.4 compound, the transition temperature, for all values of the applied field, is below 1.8 K. This observation indicates that the antiferromagnetic ordering noted for $CeNiGe_2$ [19] get shifted down in temperature, with Y-substitution. Inverse magnetic susceptibility is fitted with Curie-Weiss (CW) law (above 100 K) at 0.1 T under zero field cooling (ZFC) for all Y-substituted composition (not shown). The obtained Curie Weiss temperature ($\theta_p$) is negative which indicate the dominance of antiferromagnetic interactions. We now focus on the thermodynamic quantity - d*M*/d*T* as it is an important tool to assess the degree of freedom of delocalized spins, and to categorize QCP [16, 17]. Figure 1 (d, e and f) shows the temperature response of (-d*M*/d*T*)/*T* in field range of 0.5 – 7 T. As noted from the figure 1 (d), for Y-0.1 compound, with decreasing temperature, (-d*M*/d*T*)/*T* increases and a maximum occur around 3.6 K for 0.5 T. On further decreasing temperature a sign change from positive to negative value is noted around 3.4 K. Both, the temperature of maximum and sign change temperature shift to lower temperature with increasing field. Above 1 T, the sign change is not noted. However, with increasing field the slope of growth of (-d*M*/d*T*)/*T* increases up to 4 T and above this value of field it decreases and tends to saturate. The maximum value of slope of (-d*M*/d*T*)/*T* is around 4 T. Similar observation for sign change is noted for Y-0.2 compound up to 0.5 T and (-d*M*/d*T*)/*T* increases up to 4 T beyond of this field it decreases. For Y-0.4 compound the sign change is absent, however, maximum slope occur around 5 T. According to the relation, $\Gamma_{mag}$ = - (d*S*/d*H*)/C, where -d*S*/d*H* = -d*M*/d*T*; (-d*M*/d*T*)/*T* is related to $\Gamma_{mag}$ [16]. Hence the observed sign change in (-d*M*/d*T*)/*T* indicates to entropy accumulation and is related to magnetic frustration, as reported in CeNiGe$_2$ [19]. Hence, it can be said that magnetic frustration persists in the Y-0.1 compound. It is significantly weaker in Y-0.2 compound, whereas, it is suppressed in Y-0.4 compound.

Figure 2 (a, b, and c) shows the temperature dependence heat capacity divided by temperature (*C*/*T*) at selected fields for Y-0.1, Y-0.2, and Y-0.4 compounds. For Y-0.1 compound, two transition temperature ~ 3.1 K (weak anomaly) and ~ 2.2 K is shifted to an anomaly ~ 2.0 K for Y-0.2 compound. However, for Y-0.4 compound no transition is observed in the measured range of temperature. Under application of magnetic field, in Y-0.1 compound, two transition temperatures is shifted to the lower temperature and is suppressed below 1.8 K above 1 T, but a new anomaly appears around 4.2 K at 4 T. This new anomaly grows and shifts to lower temperature above 4 T (shown inset of figure 2 (a)). Also , for Y-0.2 compound, the transition temperature is suppressed below 1.8 K at field values of around 1 T and a new anomaly appears around 3.4 K at 5 T (shown inset of figure 2 (b)). Similar type feature has been reported for compound CeNiGe$_2$ and it is ascribed to the development of a new magnetic phase [19]. Hence, it can be said that in Y-0.1 compounds a new magnetic phase develops at high applied fields. Signatures of such features are also present in Y-0.2 compounds. Hence it can be said that Y-0.1 and Y-0.2 compounds avoids the QCP. However, no new anomaly is observed for Y-0.4 compound and $C/T$ tends to saturate above 5T at lower temperatures. Hence, due to the absence of the new anomaly in this compound, we tried to explore the possible presence of quantum criticality in Y-0.4 compound. The quantum criticality describes the collective fluctuations of matter on either side of critical field at finite temperature and result in the observation of QCP at zero temperature. In order to find the critical field of quantum criticality, we have plotted the magnetic field dependent $(-dM/dT)/T$ at lowest measured temperature of 1.8 K for Y-0.4 compound as shown in the figure 3(a). This curve is obtained from the temperature dependent $(-dM/dT)/T$ in constant fields. With increasing magnetic fields, $(-dM/dT)/T$ increases and shows a broad maximum around 5 T. In order to exactly determine the critical magnetic field ($H_C$) first derivative of the quantity $(-dM/dT)/T$ is taken (inset of figure 3 (a)). It is noted at 5 T which we assign as the critical field for quantum criticality. Here, we would like to mention that in Y-0.4 compound sign change in $(-dM/dT)/T$ is absent at $H < H_C$; as generally noted for critical transition [17]. This absence of sign change arises due to the fact that when the temperature is of the order of Kelvin, one can see broad maxima around $H_C$. This observed maxima, should grow in magnitude with decrease in temperature show a sign change for $H < H_C$ at extreme low temperature [17]. Figure 3 (b) shows the temperature response of $-dM/dT$ at the critical field for the Y-0.4 compound. The curves are fitted with the equation as:

$$-dM/dT \sim T^x \qquad \ldots\ldots\ldots (1)$$

where $x$ is the temperature exponent. As per literature reports, $-dM/dT$ varies as $T^{0.5}$ for $3d$ critical fluctuation (where $d$ is the dimensionality) and $-dM/dT$ is constant in case of $2d$ fluctuation [16, 21]. The obtained values of $x$ are $0.4 \pm 0.03$. It indicates to that parameter is nearly compatible to that observed for $3d$ fluctuations. Hence, it can be said that $3d$ fluctuations are possibly present in the critical field regime. Similar type of fluctuations has also been reported in other compounds [21, 22].

It has been reported in literature that in the quantum critical region, the correlation length ($\xi$) and correlation time ($\xi_\tau$) diverge as a function of tuning parameter $H$ or pressure ($P$). The correlation length is mathematically expressed as $\xi \sim |r|^{-\nu}$, where, $r = (H-H_C)/H_C$ and $\xi_\tau \sim \xi^z$. In the expression $\nu$ is the exponent of correlation length and $z$ is the dynamical critical exponent which depends on the dynamic of the order parameter. The free energy $F$ can be expressed using hyperscaling as [16]:

$$F(T, H) = -a(T/T_0)^{(d+z)/z} f(r/(T/T_0)^{1/\nu z}) \qquad \ldots\ldots\ldots (2)$$

where $a$ and $T_0$ are non-universal constant, $f(r/(T/T_0)^{1/\nu z})$ is a universal scaling function. Moreover, the heat capacity ($C'$) (by subtracting the lattice contribution) at $H = H_C$ and at finite $T$ is given, using a thermodynamic relation as [16, 23]:

$$C'(T, H=H_C) = -T\, \partial^2 F/\partial T^2 \qquad \ldots\ldots\ldots (3)$$

$$C'/T(T, H=H_C) \sim \gamma(0) + bT^n \qquad \ldots\ldots\ldots (4)$$

where $n = d/z - 1$ is the temperature exponent, $\gamma(0)$ is the free fitting parameter and $b$ is a constant. For Y-0.4 compound, we extract $d$ and $z$ parameter of the temperature response $C'/T$ curves at selected fields. Figure 4 (a) shows these curves, which are obtained after subtracting the phonon contribution from heat capacity up to lowest temperature using the equation:

$$C'/T = C/T - \beta T^2 \ldots\ldots\ldots (5)$$

where $\beta$ and $\gamma_p$ are the phonon and Sommerfeld coefficient respectively. $\beta$ is extracted from a linear fit of equation $C/T \sim \gamma_p + \beta T^2$ in temperature range 15 - 30 K for all fields. From figure 4 (a) it is observed that, with the increase in magnetic field, $C'/T$ increases up to 5 T, beyond which it decreases. Inset of figure 4 (a) shows the $C'/T$ curves at critical field fitted with equation (4) in the low temperature region and the obtained fitting parameters are tabulated in Table 1. The value of the exponent $z$ (extracted from $n$ and with $d = 3$) is found to be 1.9. For this compound it is observed that $d + z > 4$, implying that the critical field of this compound lies above the upper critical dimension ($d + z = 4$) in the mean field regime [16, 23]. It has been reported that, above the upper critical dimension, thermodynamic scaling of equation (2) can be affected due to presence of dangerously irrelevant variable arising out possible from low energy degree of freedom [16, 24]. However, it has been reported by authors in Refs [24, 25] that the presence of additional local critical degrees of freedom and, their coupling with the low energy degree of freedom can allow the scaling in mean field regime. This local critical mode results in continuous suppression of screening of magnetic moment towards the QCP and, this is a local phenomenon [24]. The observation of scaling in a compound suggests the relevance of coupling of the local critical mode to the fluctuation resulting in the observation of possibly quantum criticality.

Hence, in order to perceive the quantum criticality in mean field regime, we do the $H/T$ scaling. Equation (2) is rearranged by choosing $T_0 = \ell^z T$, we obtain [17, 23]:

$$F(T, H) = \ell^{-(d+z)} F(\ell^z T, \ell^{\beta\delta/\nu} H) \qquad \ldots\ldots (6)$$

where ℓ is an arbitrary scale factor, δ and β are the order parameters depending upon field and concentration respectively. Using equation (6) with cut-off energy $k_B T_0$ (where $k_B$ is the Boltzmann constant), equation (3) is obtained as

$$C'(T, H)/T^{d/z} = \Psi(H/T^{\beta\delta/\nu z}) \quad \ldots\ldots (7)$$

where $d/z$ and $\beta\delta/\nu z$ are scaling factor coupled with the parameters, magnetic field and temperature and $\Psi(H/T^{\beta\delta/\nu z})$ is scaling function. Figure 4 (b) shows the scaling of heat capacity, $\Delta C'/T^{d/z}$ ($\Delta C = C'(T, H) - C'(T, H_C)$) versus $\Delta H/T^{\beta\delta/\nu z}$ ($\Delta H = H - H_C$) in different fields for Y-0.4 compound. In this curve the heat capacity at critical field is subtracted to exclude the non-critical contributions. A best scaling is observed for $(d/z, \beta\delta/\nu z) = (1.6, 2.5 \pm 0.04)$ ($d/z$ is estimated from the equation (4)). The scaling of heat capacity indicates to presence of quantum criticality. Similar form of scaling has already been reported in other compounds exhibiting quantum criticality [23, 26-27].

In order to determine the critical scaling parameter δ, we also do the scaling of magnetization curves of these compounds. In terms of free energy, $M(T, H) = -\partial F/\partial H$. Using equation (6), the magnetization is expressed as

$$M(T, H)/T^{\beta/\nu z} = \Phi(H/T^{\beta\delta/\nu z}) \quad \ldots\ldots (8)$$

where $\beta/\nu z$ is scaling factor coupled with the parameters, magnetic field and temperature and $\Phi(H/T^{\beta\delta/\nu z})$ is scaling function. Figure 4 (c) shows a scaling of $M(T, H)/T^{\beta/\nu z}$ versus $H/T^{\beta\delta/\nu z}$ in field ranges 5.5 - 7 T (since the maximum range of MPMS system is 7 T). The best scaling is achieved (above the critical field) for $\beta\delta/\nu z = 2.5 \pm 0.04$ and $\beta/\nu z = 1.56 \pm 0.02$. Using these parameters, the obtained value of δ is 1.6. It is noted that δ deviates from the mean field value of 3. This mean field value determines the size of fluctuation across the critical field [23, 28]. The obtained small value of δ suggests the presence of the existence of unusual fluctuation. These unusual fluctuations indicate a deviation from the typical $3d$ critical fluctuation and probably arise due to disorder effect because of Y-substitution at the Ce-site [29]. Hence it can be said that anomalous quantum criticality is observed in Y-0.4 compound. Such features are not uncommon and have been reported in other compounds [30].

## 4. Summary:

In this work, an investigation is carried out to check the presence of magnetic field induced quantum criticality in Ce-site diluted heavy fermion $CeNiGe_2$. Results of temperature response of thermodynamic quantity -d$M$/d$T$ reveal that magnetic frustration persists in Y-0.1 compound, which weakens in Y-0.2 compound and is suppressed in Y-0.4 compound. For Y-0.1 and Y-0.2 compounds, from heat capacity studies, it is observed that a new magnetic phase develops at high field, as a result of which quantum criticality is not noted in these two compounds. No such phase is observed in Y-0.4 compound and it obeys $H/T$ scaling of quantum criticality above critical field. However, the obtained critical scaling parameter deviates from the mean field value due to disorder effect induced by Y-substitution. This observation suggests the presence of unusual fluctuations and anomalous quantum criticality in Y-0.4 compound. Hence this work gives evidence of field induced anomalous quantum criticality in Y-0.4 compound via dilution of Ce-site, which is impeded by the development of new magnetic phase at high fields in CeNiGe$_2$. Further low temperature (in milli-Kelvin range) measurement is needed in future to determine the existence of QCP in these compounds.

**Acknowledgements**

The authors acknowledge experimental facilities of Advanced Material Research Centre (AMRC), IIT Mandi. Financial support from IIT Mandi is also acknowledged.

**References**

[1] G. R. Stewart, Rev. Mod. Phys. **56** (1984) 755; Rev. Mod. Phys. **73**, (2001) 797

[2] N. D. Mathur *et al*., Nature (London) **394** (1998) 39

[3] H. v. Lohneysen *et al*., Rev. Mod. Phys. **79** (2007) 1015

[4] P. Gegenwart *et al*., Nature Phys. **4** (2008) 186

[5] C. Petkovic *et al*., J. Phys.: Condens. Matter **13** (2001) L337

[6] A. Yeh *et al*., Nature (London) **419** (2002) 459; H. v. Lohneysen *et al*., Phys. Rev. Lett. **72** (1994) 3262; P. Gegenwart *et al*., Phys. Rev. Lett. **89** (2002) 056402

[7] Hulliger F. Handbook Phys. Chem. Rare Earths **4** (1979) 153-236; R. Siemen and B. R. Cooper, Phys. Rev. B **19** (1979) 2645

[8] V. F. Correa *et al*., J. Phys.: Condens. Matter **28** (2016) 346003; W. G. C. Oropesa *et al*., J. Phys.: Condens. Matter **30** (2018) 295803

[9] A. J. Millis, Phys. Rev. B. **48** (1993) 7183

[10] T. Moriya and T. Takimoto, J. Phys. Soc. Jpn. **64** (1995) 960

[11] G. Oomi et al. Transport and Thermal Properties of *f*-electron Systems. Plenum Press, New

York, (1993)

[12] V. K. Pecharsky and K. A. Gschneidner, Jr. Phys. Rev. B **43**, (1991) 10906; C. Geibel *et al.*, J. Magn. Magn. Mater. **108** (1992) 207; P. Schobinger-Papamantellos *et al.*, J. Magn. Magn. Mater. **125** (1993) 151; A.P.Pikul *et al.*, J. Magn. Magn. Mater. **16** (2004) 6119

[13] M. H. Jung *et al.*, Phys. Rev. B **66** (2002) 054420

[14] A. T. Holmes *et al.*, Phys. Rev. B **85** (2012) 033101.

[15] D. Y. Kim *et al.*, J. Phys.: Condens. Matter **16** (2004) 8323

[16] L. Zhu *et al.*, Phys. Rev. Lett **91** (2003) 066404

[17] M. Garst and A. Rosch, Phy. Rev. B **72** (2005) 205129

[18] A. Sakai *et al.*, Phys. Rev. B **94** (2016) 220405.

[19] Karan Singh and K. Mukherjee, Phys. Lett. A **381** (2017) 3236

[21] Karan Singh and K. Mukherjee, Philos. Mag. DOI:10.1080/14786435.2018.1538577 (2018)

[21] J. G. Donath *et al.*, Phys. Rev. Lett **100** (2008) 136401

[22] R. Kuchler *et al.*, Phys. Rev. Lett **96** (2006) 256403

[23] C. L. Huang *et al.*, Nat. commu. **6** (2015) 8188

[24] Q. Si, S. Rabello *et al.*, Nature (London) **413** (2001) 804;

[25] D. R. Grempel and Q. Si, Phys. Rev. Lett **91** (2003) 026401; S. Sachdev, Phys. Rev. B **55** (1997) 142; L. Mistura, J. Chem. Phys. **57** (1972) 2306

[26] A. Bianchi *et al.*, Phys. Rev. Lett **91** (2003) 257001

[27] K. Heuser *et al.*, Phys. Rev. B **58** (1998) 15959

[28] J. A. Hertz, Phys. Rev. B **14** (2000) 1165

[29] C. H. Wang *et al.*, J. Phys.: Condens. Matter **27** (2015) 015602

[30] Y. Itoh *et al.*, J. Phys. Soc. Jpn **77** (2008) 12370

**Table 1:** Parameters ($\gamma$ (0), b and $n$) obtained from the $C'/T$ curves fitting with equation (4) in the low temperature region at critical fields. The values of $\beta\delta/\nu z$, $\beta/\nu z$ and $\delta$ are obtained from the scaling of heat capacity (equation (7)) and magnetization (equation (8)) for the Y-0.2 and Y-0.4 compounds

| Compound | $\gamma$ (0) (J/mol-K$^2$) | $b$ (J/mol-K$^{2-n}$) | $n = d/z-1$ | $\beta\delta/\nu z$ | $\beta/\nu z$ | $\delta$ |
|---|---|---|---|---|---|---|
| Y-0.4 | 0.711 | -0.187 | 0.6± 0.01 | 2.5 ± 0.04 | 1.56 ± 0.03 | 1.6 |

**Figures:**

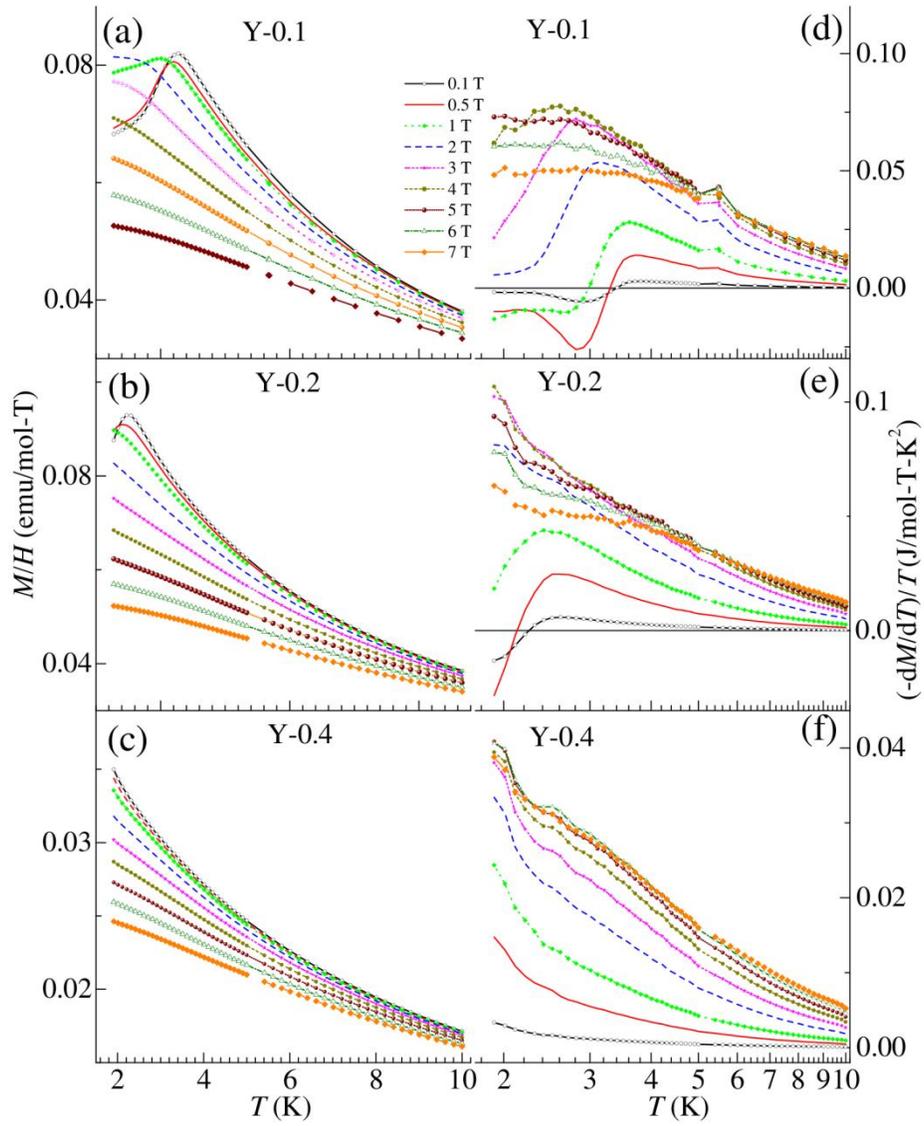

Figure 1 (a), (b), and (c): Temperature (*T*) response DC susceptibility (*M*/*H*) under zero field cooling condition at different fields for Y-0.1, Y-0.2, and Y-0.4 compounds respectively. (d), (e), and (f): Temperature dependent (-d*M*/d*T*)/*T* divided by temperature for the same fields.

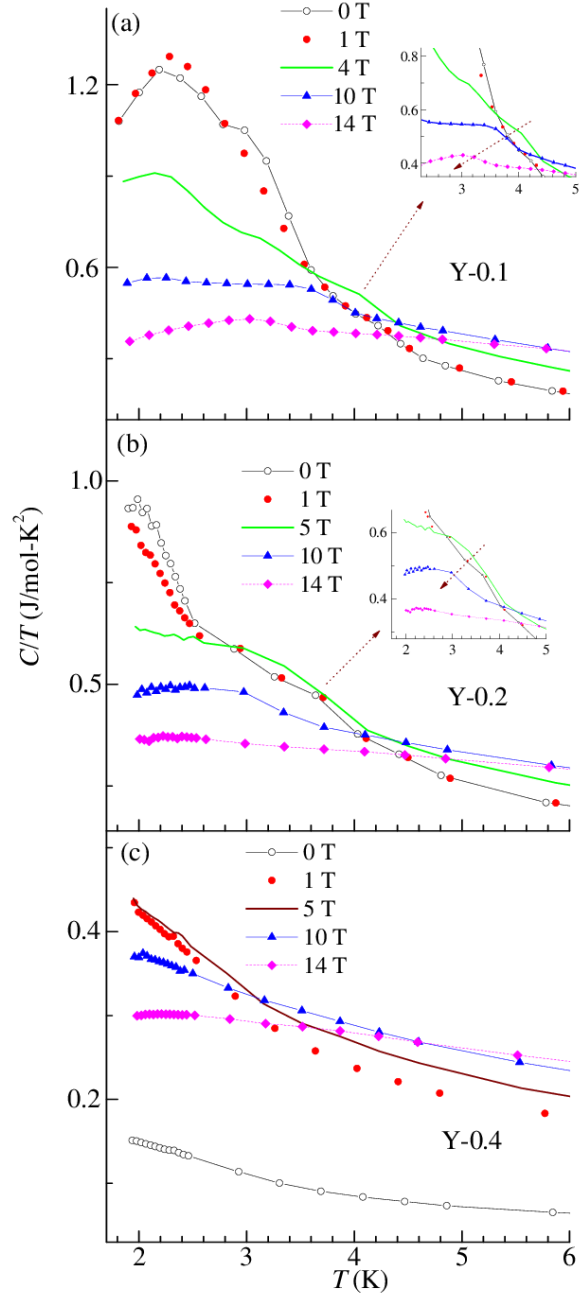

Figure 2 (a), (b), and (c): Temperature (*T*) response heat capacity divided by temperature (*C*/*T*) at selected fields for Y-0.1, Y-0.2, and Y-0.4 respectively. Insets of (a) and (b): Magnified plot in the temperature range of 1.8 – 5 K. Arrow indicate the position of shifting of transition temperature. The curves for 0 T is added from Ref [20].

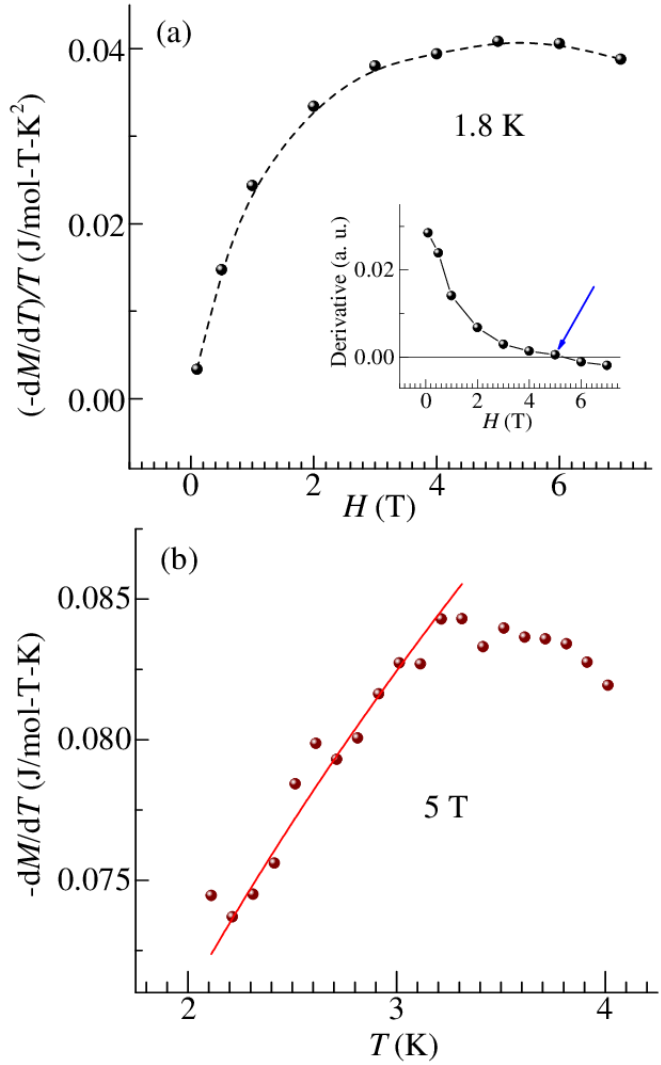

Figure 3 (a): Magnetic field (*H*) response thermodynamic quantity -d*M*/d*T* divided by temperature at 1.8 K for Y-0.4 compound. Inset: derivative of (-d*M*/d*T*)/*T*. Arrow indicate the point of crossover. (b): -d*M*/d*T* ~ $T^x$ curve fitting at 5 T for Y-0.4 compound.

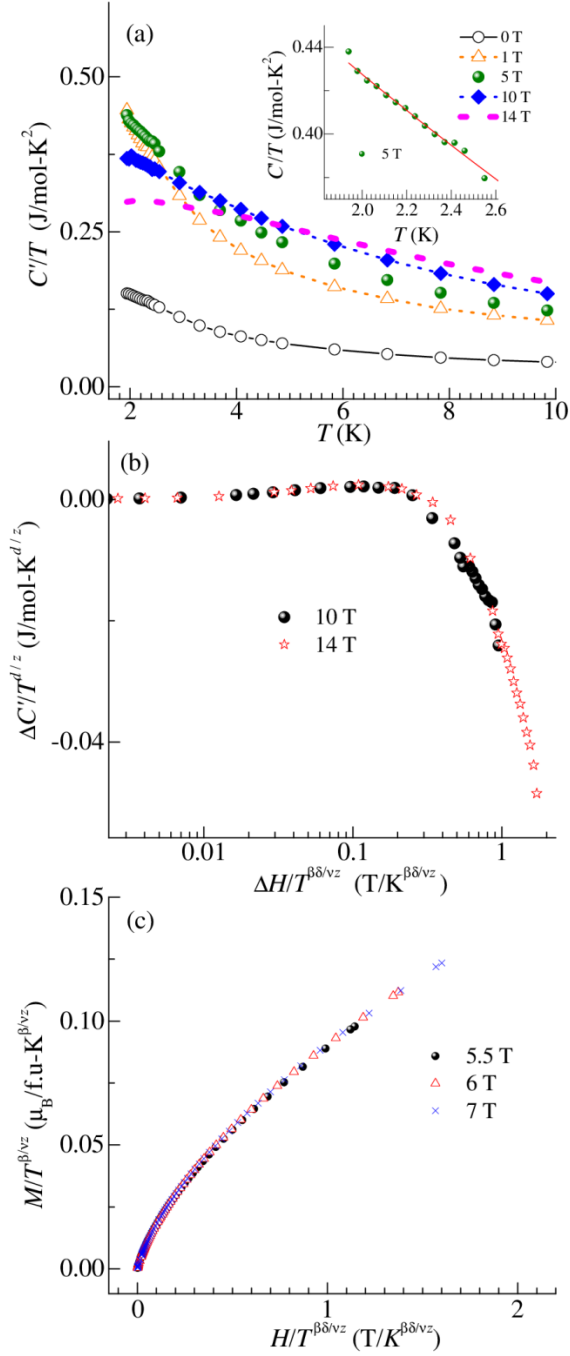

Figure 4 (a): Temperature response of $C'/T$ for Y-0.4 compound at selected magnetic fields. Inset: temperature dependence of $C'/T$ at field at 5 T. Solid red lines are the fit of equation $C'/T \sim \gamma(0) + a\,T^n$. (b): Scaling of field dependence heat capacity (heat capacity contribution at critical field subtracted) for Y-0.4 compound. (c): Scaling of $M(T, H)/T^{\beta/\nu z}$ versus $H/T^{\beta\delta/\nu z}$ for Y-0.4 compound.

13